\documentclass{article}
\usepackage{times,geometry,enumerate}
\usepackage[dvips]{graphicx}
\usepackage{subfigure}
\usepackage{hyperref}

\providecommand{\keywords}[1]{\textbf{Keywords:} #1}
\begin{document}
\author{Cristina Blaga\\Faculty of Mathematics and Computer Science \\Babe\c{s}-Bolyai University, Cluj-Napoca, Romania}
\title{Stability in sense of Lyapunov of circular orbits in Manev potential}
\date{}
\maketitle
\begin{abstract}
In this article we consider the motion of two bodies under the action of a Manev central force. We obtain the radius of the circular orbit and analyze its stability in sense of Lyapunov. Drawn on the first integrals of angular momentum and energy, we build a positive definite function which satisfies the Lyapunov's theorem of stability. The existence of the Lyapunov function prove that the circular orbits in Manev two body problem are stable at any perturbation. In the end we compare these results with those valid for the circular orbits in the Newtonian gravitational field.      
\end{abstract}

\keywords{Manev potential, stability in sense of Lyapunov, circular orbits}
%Further we determine the Lagrangian of the motion. Drawn on the existence of a cyclic coordinate for the Lagrangian, we compute the Routh function and obtain a criteria for the stability of the steady circular motion.
\section{Introduction}

In the nineteen twenties, Georgi Manev (Maneff in German and French spelling) published several papers (\cite{m25}, \cite{m29}, \cite{m30}) in which he proposed a nonrelativistic gravitational law, able to explain certain dynamical phenomena observed in the solar system, not explained in the frame of classical mechanics. Manev obtained his model as a consequence of Max Planck's action-reaction principle in~\cite{m25} and considered it as a substitute to general relativity. In his papers he noticed that this model provides a good explanation to the observed motion in the solar system. Manev emphasized that under a particle of mass $m_2$, which is moving in the static gravitational field due to mass $m_1$ $(m_1>m_2)$, acts a central force of the form     
\begin{equation}\label{fM}
F(r)=-\frac{G m_1 m_2}{r^2} \left( 1 + \frac{3 G (m_1+m_2)}{c^2 r} \right)
\end{equation}
where $r$ is the distance from the particle $m_2$ to the center of mass of $m_1$, $G$ the Newtonian gravitational constant and $c$ the velocity of light. The force~(\ref{fM}) - known as Manev force - differs from the Newtonian force through the additional term inverse proportional with $r^{3}$. Under the action of Manev force the mass $m_2$ describes a precessional ellipse, with $m_1$, assimilated to a point mass, placed in one focus of the ellipse. Around the point in which $m_1$ is situated, the apsidal line is rotating and so perihelion advance of Mercury could be qualitatively explained. Sir Isaac Newton considered a similar force in his book \emph{Philosophiae Naturalis Principia Mathematica} issued in 1687. He proved that if to the Newtonian gravitational force is added the term $\mu/r^3$ (where $\mu>0$ and $r$ the distance between the particles), the particle describes a precessional ellipse (see \cite{n99}, Book I, Section IX, Proposition XLIV, Theorem XIV). At that time, Newton was looking for an explanation to the observed motion of the perigee of Moon. There were other attempts to use forces similar to Manev force to explain certain dynamical phenomena observed in the solar system, unexplained in the classical mechanics. A comprehensive list of such historical works is given in Diacu \emph{et.al.}~\cite{d95}. 

After several decades in which Manev model was forgotten, it was brought to light by Diacu~\cite{d93}. In nineteen nineties the motion under Manev force was analysed in order to model different phenomena from celestial mechanics, stellar dynamics or from astrophysics. A survey of the papers devoted to the applications of Manev gravitational field could be found in Haranas and Mioc~(\cite{hm09}).   

The investigation of the motion in Manev's gravitational field revealed that in the solar system it gives a theoretical approximation as good as general relativity, building a bridge between classical mechanics and general relativity. An investigation of the ways in which modified Manev potential could model certain general relativistic results in the frame of Newtonian mechanics was done by Ivanov and Prodanov~\cite{ip05}. They analyzed the circular orbits around a rotating and non-rotating massive body or in a parametrized post-newtonian metric.   

The two body problem in Manev potential was considered by several authors. The analytic solution of the problem and its behavior near collision was investigated by Diacu \emph{et.al}~\cite{d95}. The solution of the regularized equations of motion for different initial conditions and hypothesis about regularization was obtained by Mioc and Stoica in~\cite{ms95a} and~\cite{ms95b}.  

In this paper we consider the circular orbits in the two-body problem in Manev potential. In space dynamics the circular orbits are met often. The spacecrafts exploring the Universe are moving sometimes on circular orbits and also the transfer between the bodies from the solar system is done sometimes on segments of circular paths (see~\cite{r04}).
% At the beginning of the space era, the orbiters meant to study the Moon, Venus or Mercury described segments of ellipses with small eccentricities. In that case the near circular orbits was preferred, because if the orbiter would have moved on an elongated orbit it could step out from the sphere of influence of the surveyed body. Even now, the orbiters meant to survey the small bodies from the solar system, like dwarf planets, asteroids or comet nuclei, are designed to revolve around the body on near circular orbits to avoid the exit from the sphere of influence of the body. After 2000, in the vicinity of Earth were launched satellites in low, near circular polar orbits to make precise measurements of Earth's gravity field (GRACE - twin satellites),
% - \url{http://www.csr.utexas.edu/grace/} %\verb|http://www.csr.utexas.edu/grace/|) 
%Earth's magnetic field (SWARM - a constelation of three satellites)  
%- \url{http://www.esa.int/Our_Activities/Observing_the_Earth/The_Living_Planet_Programme/Earth_Explorers/Swarm} . %\verb|http://www.esa.int/Our_Activities/Observing_the_Earth/The_Living_Planet_Programme/Earth_Explorers/Swarm|).           
%or both gravity and magnetic field (CHAMP mission between 2000 and 2010).
%http://www.gfz-potsdam.de/en/research/organizational-units/departments/department-2/earths-magnetic-field/infrastructure/champ/
The geostationary satellites have also near circular orbits.

We study here the stability in the sense of Lyapunov of the circular orbits in Manev potential. In the section $2$ we obtain the equations of motion of a body under the action of Manev central force~(\ref{fM}) and determine the radius of the circular orbit. In section $3$, we derive the Lagrangian and obtain the first integrals of motion. Drawn on the angular momentum and energy integral, we build a positive defined function which fulfills the conditions from the Lyapunov's stability theorem. Using this function on the ground of the Lyapunov's theorem of stability we reach the conclusion that the circular orbit is stable at any perturbation. In the last section we compare characteristics of the circular orbit in Manev potential with those from the Newtonian gravitational field.  

\section{Circular orbits in Manev's potential}

\subsection{Equations of motion}

Let us consider the motion of two interacting bodies of masses $m_1$ and $m_2$ ($m_1>m_2$) under a Manev force (\ref{fM}). We assume that bodies are assimilated to point masses located in their center of mass and the Manev force is an attracting force, acting in the line joining the bodies. If the position vectors of the bodies are $\vec{r_1}$ and $\vec{r_2}$ respectively, the equations of motion are:
\begin{eqnarray}
m_1 \ddot{\vec{r}}_1 &=& \vec{F}_{12} \label{m1}\\
m_2 \ddot{\vec{r}}_2 &=& \vec{F}_{21} \label{m2}
\end{eqnarray}
where $$\vec{F}_{12}=-\vec{F}_{21}=-F \, \frac{\vec{r}}{r}$$ and $\vec{r}=\vec{r}_2 - \vec{r}_1$ is the position vector of $m_2$ relative to $m_1$. After several straightforward transformations  
%Multiplying equ.~(\ref{m1}) with $m_2$, equ.~(\ref{m2}) with $m_1$ and subtracting from the first the second equation, 
we get the equation of relative motion of $m_2$ in respect with $m_1$
\begin{equation}
\mu \ddot{\vec{r}}=F(r) \frac{\vec{r}}{r} \label{mr} \,,
\end{equation} 
where $\mu = m_1 m_2/(m_1 + m_2)$ is the reduced mass.

During motion, the force acting on $m_2$ is directed to the center of mass of $m_1$, it is a central force. It depends only on the radial coordinate $r$, therefore we can determine a scalar function $V(r)$ from which Manev force is derived ($F= -d V(r)/d r$). This function known as Manev potential and is given by  
\begin{equation}\label{VM}
V(r)=-\frac{G m_1 m_2}{r} \left( 1 + \frac{3 G (m_1+m_2)}{2 c^2 r} \right)\,,
\end{equation}
where quantities $m_1$, $m_2$, $G$, $c$ and $r$ have the same meaning as in~(\ref{fM}).

If we multiply vectorial the equation of relative motion~(\ref{mr}) with position vector $\vec{r}$, we get the angular momentum integral 
\begin{equation}\label{iav}
	\vec{r} \times \mu \vec{\dot{r}} = \vec{C} 
\end{equation}
where the vector $\vec{C}$ is constant, perpendicular on the plane in which the motion took place. Like in the Newtonian case, the motion is restricted to the plane determined by the initial position of $m_1$, the initial position vector of $m_2$ relative to $m_1$ and its initial velocity in respect to $m_1$. 

If we multiply scalar~(\ref{mr}) with the relative velocity of $m_2$ in respect with $m_1$, $\vec{v}=\dot{\vec{r}}$, after some algebra we get the energy integral
\begin{equation}\label{ie}
\frac{\mu}{2} v^2 + V(r) = h 
\end{equation}
where $v$ is the length of velocity and $h$ the total energy. The relation~(\ref{ie}) stats that the total energy is conserved during the motion. 

%Let us introduce the Lagrangian 
%\begin{equation}\label{L}
% L=T+V(r)
%\end{equation}
%where $T=\mu v^2 /2 $ is the kinetic energy. From~(\ref{ie}) we get that the Lagrangian is constant.

The motion of $m_2$ is planar, thus further we will analyze the motion in the orbital plane. In polar coordinates ($r$,$\varphi$) the acceleration of $m_2$ is 
\begin{equation}
\ddot{\vec{r}}=(\ddot{r}-r {\dot{\varphi}}^2) \vec{e}_r + (r \ddot{\varphi} + 2 \dot{r} \dot{\varphi}) \vec{e}_{\varphi}\,\,,
\end{equation}
where $\vec{e}_r$ and $\vec{e}_{\varphi}$ are the unit vectors on the radial and polar axis respectively. After some straightforward algebra, the projections of the equation of relative motion~(\ref{mr}) on the radial and polar axis become: 
\begin{eqnarray}
\ddot{r}-r {\dot{\varphi}}^2 &=& - \frac{\gamma}{r^2} \left( 1 + \frac{3 \gamma}{c^2 r} \right) \label{mrer}\\
\frac{1}{r} \frac{d}{d t} (r^2 \dot{\varphi}) &=& 0 \label{ia}
\end{eqnarray}        
where $\gamma=G(m_1+m_2)$ is the gravitational parameter. From~(\ref{ia}) $r^2 \dot{\varphi}=C$, with $C$ a constant. The relation ~(\ref{ia}) is equivalent with integral of angular momentum~(\ref{iav}), because the length of angular momentum is $|\vec{C}| = \mu r^2 \dot{\varphi}$.   

\subsection{Existence of circular orbits}

In polar coordinates, the circular orbits are given by $r=r_0=\mbox{constant}$, where the constant $r_0$ is the radius of the circular orbit. From~(\ref{ia}) $\dot{\varphi}=C/r_0^2$, the angular velocity is constant and the circular motion is uniform, like in the Newtonian case. On a circular orbit $\dot{r}=\ddot{r}=0$ and from~(\ref{mrer}) we get
\begin{equation}\label{r0}
r_0=\frac{C^2}{\gamma}\left( 1- \frac{3 \gamma^2}{C^2 c^2} \right)\,.
\end{equation}
We observe that if $C^2>3 \gamma^2/c^2$, then the positive real number $r_0$ given by~(\ref{r0}) is the radius of the circular orbit in the Manev two-body problem. In the classical Newtonian two-body problem the radius of the circular orbit is $r_{0}=C^2/\gamma$ (see~\cite{g80}). Comparing the radius of the circular orbit in the Manev and Newtonian potential, for a given angular momentum, we note that in the Manev case the radius of the circular orbit is smaller then in the Newtonian case. For a given angular momentum constant $C$, the difference between the radius of the circle described in the Manev potential~(\ref{r0}) and the corresponding circular orbit from the Kepler problem is
\begin{equation}\label{Dr0}
\Delta r_0=r_{0N}-r_{0M} = \frac{3 \gamma^2}{c^2} \,,
\end{equation}        
where $r_{0N}$ and $r_{0M}$ are the radius of the circular orbits in Newtonian and Manev potential, respectively. We emphasize that the difference~(\ref{Dr0}) does not depend on $C$, although the radius of the circular orbits are dependent on $C$. The difference depends on the ratio between the gravitational parameter and the speed of light.  
 
\section{Stability of the circular orbit}

Let us consider the stability of the circular orbit in Manev two-body problem in respect with the variables $r$, $\theta$, $\varphi$ and their time derivatives $\dot{r}$, $\dot{\theta}$, $\dot{\varphi}$, where ($r$, $\theta$, $\varphi$) are the spherical coordinates, $r$ is the radial coordinate, $\theta$ the latitude and $\varphi$ the longitude. We assume that the circular orbit lies in the equatorial plane.    

The Lagrangian for the Manev two-body is 
\begin{equation}\label{L}
L=T-V(r) 
\end{equation}
where the kinetic energy $T$ in spherical coordinates is 
\begin{equation}
T=\frac{\mu}{2}\left( \dot{r}^2 + r^2 \dot{\theta}^2 + r^2 \cos \theta  \dot{\varphi}^2 \right) 
\end{equation}
and the potential $V(r)$ from~(\ref{VM}) in terms of the reduced mass $\mu$ and the gravitational parameter $\gamma$ is  
\begin{equation}
V(r)= -\frac{\mu \gamma}{r} \left( 1 + \frac{3 \gamma}{2 c^2 r} \right)\,.
\end{equation}

We observe that $\varphi$ does not appear explicitly in the Lagrangian~(\ref{L}), it is a cyclic coordinate. Therefore from the Lagrange equation for $\varphi$ we get the following integral of motion
\begin{equation}\label{F2}
r^2 \cos ^2 \theta \dot{\varphi} = b \,,
\end{equation}   
with $b$ a constant.

The perturbed motion is given by $x_i$, $i=\overline{1,5}$, with
\begin{equation}
r = r_0 +x_1 \,, \quad \dot{r} = x_2 \,, \quad \theta = x_3 \, , \quad \dot{\theta} = x_4 \, , \quad \dot{\varphi}=\dot{\varphi_0} + x_5 \,,
\end{equation}
where $\dot{\varphi_0}=C/r_0^2$ and $r_0$ is given by~(\ref{r0}).   

According to the stability theorem in Lyapunov sense, we are looking for a function $\mathcal{F}(\mathbf{x})$, with $\mathbf{x}=(x_1,x_2,x_3,x_4,x_5)$, which is positive definite in a vicinity of the unperturbed motion $\mathbf{x}=\mathbf{0}$ (see~\cite{h67}). Usually, the first integrals of motion are used to build the Lyapunov function. Therefore, we consider the energy integral~(\ref{ie}), divided by $\mu/2$, for the perturbed motion
\begin{equation}\label{F1p}
F_1(\mathbf{x})=x_2^2+(r_0+x_1)^2 x_4^2 +(r_0+x_1)^2 (\dot{\varphi}+x_5)^2 \cos^2 x_3 - \frac{2 \gamma}{r_0+x_1} - \frac{3 \gamma^2}{c^2 (r_0+x_1)^2} 
\end{equation}
and the integral of motion~(\ref{F2}) for the perturbed motion
\begin{equation}\label{F2p}
F_2(\mathbf{x})=(r_0+x_1)^2 (\dot{\varphi}+x_5) \cos^2 x_3 \,.
\end{equation}
These functions do not have definite sign in respect with $x_i$, $i=\overline{1,5}$, thus we consider the function
\begin{equation}\label{F}
\mathcal{F} (\mathbf{x}) = F_1(\mathbf{x})-F_1(\mathbf{0})+\lambda [F_2(\mathbf{x})-F_2(\mathbf{0})]+\nu [F_2^2(\mathbf{x})-F_2^2(\mathbf{0})]\,, 
\end{equation}    
where $\mathbf{0}=(r_0,\varphi_0,0,\dot{\varphi_0},0)$ denotes the circular unperturbed motion and $\lambda$, $\nu$ are two constants. Further, we examine whether there are real numbers $\lambda$ and $\nu$, so that $\mathcal{F} (\mathbf{x})$ is a positive definite function in a neighborhood of the circular unperturbed motion, given by $\mathbf{0}=(r_0,\varphi_0,0,\dot{\varphi_0},0)$.

Developing in Taylor series $\cos^2 x_3$ in the vicinity of $0$ and $2 \gamma/r^2+3 \gamma^2/(c^2 r^2)$  in the neighborhood of $r_0$, neglecting the terms of order 3 and higher and keeping in mind that from~(\ref{mrer}) for a circular orbit 
\begin{equation}\label{pp0}
r_0 \dot{\varphi_0}^2= \frac{\gamma}{r_0^2} \left(1+\frac{3 \gamma}{c^2 r_0} \right)
\end{equation}
we get   
\begin{equation}\label{F1x}
F_1(\mathbf{x})-F_1(\mathbf{0}) = 4 r_0 \dot{\varphi}_0^2 x_1 + 2 r_0^2 \dot{\varphi}_0 x_5 - \left( \dot{\varphi}_0^2 + \frac{3 \gamma^2}{c^2 r_0 ^4}\right) x_1^2 + x_2^2 - r_0^2 \dot{\varphi}_0^2 x_3^2 + r_0 \left( x_4^2 + x_5^2 \right) + 4 r_0 \dot{\varphi}_0 x_1 x_5 
\end{equation}   
\begin{equation}\label{F2x}
F_2(\mathbf{x})-F_2(\mathbf{0}) = 2 r_0 \dot{\varphi}_0 x_1 + r_0^2 x_5 + \dot{\varphi}_0 x_1^2 - r_0^2 \dot{\varphi}_0 x_3^2 + 2 r_0 x_1 x_5
\end{equation}
and 
\begin{equation}\label{F22x}
F_2(\mathbf{x})^2-F_2(\mathbf{0})^2 = 4 r_0^3 \dot{\varphi}_0^2 x_1 + 2 r_0^4 \dot{\varphi}_0 x_5 + 6 r_0^2 \dot{\varphi}_0^2 x_1^2 - 2 r_0^4 \dot{\varphi}_0^2 x_3^2 + 8 r_0^3 \dot{\varphi}_0 x_1 x_5 + r_0^4 x_5^2\,.
\end{equation}
After replacing~(\ref{F1x}),~(\ref{F2x}) and~(\ref{F22x}) in~(\ref{F}), we observe that the necessary and sufficient condition for $\mathcal{F}$ to have an extremum in $x_i=0$, $i=\overline{1,5}$ is
\begin{equation}\label{l}
\lambda = - 2 \dot{\varphi}_0 \left( 1 + r_0^2 \nu \right) \,.
\end{equation}     
We substitute $\lambda$ from~(\ref{l}) in~(\ref{F}) and write $\mathcal{F} (\mathbf{x})$ as a sum of two functions 
\begin{equation}
\mathcal{F} (\mathbf{x}) = \mathcal{F}_1 (x_2,x_3,x_4) + \mathcal{F}_2 (x_1,x_5)
\end{equation}  
with 
\begin{equation}
\mathcal{F}_1 (x_2,x_3,x_4) = x_2^2 + r_0^2 \dot{\varphi}^2 x_3^2 + r_0^2 x_4^2
\end{equation}
a positive definite function in respect with $x_2$, $x_3$, $x_4$ and 
\begin{equation}
\mathcal{F}_2 (x_1,x_5) = c_{11} x_1^2 + 2 c_{12} x_1 x_5 + c_{22} x_5 \,,
\end{equation}
with 
\begin{equation} 
c_{11}=4 \dot{\varphi}_0^2 \left(r_0^2 \nu - 1 \right) + \frac{\gamma}{r_0^3}  , \quad  c_{12}=c_{21}=2 r_0^3 \dot{\varphi}_0 \nu, \quad  c_{22}= r_0^2 \left( 1 + \nu r_0^2 \right), 
\end{equation}
a quadratic form in $x_1$ and $x_5$. $\mathcal{F}_2 (x_1,x_5)$ is positive definite if and only if all principal minors are positive. In other words, if $c_{11}>0$ and $d=c_{11} c_{22}-c_{12}^2>0$. After some algebra, using~(\ref{pp0}), these two inequalities lead us to 
\begin{equation}\label{nu}
\nu> \max \left\lbrace \frac{3 \gamma \left(c^2 r_0 +4 \gamma \right)}{c^2 r_0^6 \dot{\varphi}_0^2}, \frac{3 \left(c^2 r_0 + 4 \gamma \right)}{r_0^3 c^2} \right\rbrace \,.
\end{equation}     
It is easy to check that~(\ref{nu}) is fulfilled if $\nu>3(c^2 r_0 + 4 \gamma)/(r_0^3 c^2)$. Thus $\mathcal{F}_2 (x_1,x_5)$ is a positive definite function if $\nu$ satisfies~(\ref{nu}). Further, $\mathcal{F} (\mathbf{x})$ is a positive definite function and we reach the conclusion that the circular orbit~(\ref{r0}) is stable in the sense of Lyapunov. This result generalize the Lyapunov stability of circular orbits in Newtonian gravitational field (\cite{d76},~\cite{g80}).  

\section{Conclusions}

The aim of the study was to examine the stability in the sense of Lyapunov of circular orbits in Manev two body problem. The central force acting on the bodies differs from the Newtonian force through the term inverse proportional with $r^3$, therefore we will compare the results of this study with those from the classical Newtonian two-body problem (\cite{d76},\cite{g80}). 

For two given bodies of masses $m_1$ and $m_2$ ($m_1>m_2$) under the action of the central Manev force \ref{fM} we obtained that the radius of the circular orbit described by $m_2$ depends on the angular momentum constant $C$, like in the Newtonian case. For a given $C$, the circle described in Manev's potential has a smaller radius. The quantity with which the radius of the circular orbit from the Manev case is smaller than the radius from the Newtonian case does not depend on $C$. 

From $C=r_0^2 \dot{\varphi}_0$ we get that the angular velocity on the circular orbit is constant, the motion is uniform. Using the relation between the radius of the circular orbit in Manev and Newtonian case, we get that for a given $C$, between the corresponding angular velocities exists the inequality $\dot{\varphi}_{0M}=C/r_{0M}^2>C/r_{0N}^2=\dot{\varphi}_{0N}$. In other words, for a given angular momentum, the secondary body in Manev two body problem has a higher angular velocity then in Kepler problem. In addition, the period of the circular motion in Manev two body problem is smaller then in Newtonian case, because $T_{M}=2 \pi/\dot{\varphi}_{0M}< 2 \pi/ \dot{\varphi}_{0N} = T_{N}$.

\begin{table}[h]
	\centering
	\begin{tabular}{|l|c|c|}
		\hline  &Manev&Newton\\ \hline
		Force $F(r)$&$-\frac{G m_1 m_2}{r^2} \left( 1 + \frac{3 G (m_1+m_2)}{c^2 r} \right)$&$ - \frac{G m_1 m_2}{r^2} $ \\ \hline
		Potential $V(r)$&$ -\frac{G m_1 m_2}{r} \left( 1 + \frac{3 G (m_1+m_2)}{2 c^2 r} \right)$ &$ - \frac{G m_1 m_2}{r}$ \\ \hline
		Radius of circular orbit $r_0$&$\frac{C^2}{\gamma}  \left( 1 - \frac{3 \gamma^2}{C^2 c^2} \right) $ &$\frac{C^2}{\gamma}$ \\ \hline
		Relation between&& \\ 
		$\nu$ and $\lambda$&$\lambda = - 2 \dot{\varphi}_0 \left( 1 + r_0^2 \nu \right)$& $\lambda = - 2 \dot{\varphi}_0 \left( 1 + r_0^2 \nu \right)$ \\ \hline 
		Necessary and sufficient && \\
		condition for the stability &$\nu > \frac{3 \left( c^2 r_0 + 4 \gamma \right)}{c^2 r_0^3}$& $\nu > \frac{3}{r_0^2}$  \\   
		of the circular orbit && \\ \hline
	\end{tabular}
	\caption{Circular orbits in Manev and Newtonian potential. The following notations were used $\mu=m_1 m_2/(m_1+m_2)$, $\gamma=G(m_1+m_2)$ and $c$ the speed of light. The relation between the constants $\nu$ and $\lambda$ and the necessary and sufficient condition for the stability of the circular orbit in the Newtonian potential are from~\cite{d76}.}
	\label{tabelul1}
\end{table}

In table~(\ref{tabelul1}) we gathered the results about the stability in sense of Lyapunov of circular orbit in Manev and Newtonian two-body problem. The radius of the  orbits in these two cases are different for a given $C$, but the analytical expression for the relation between the unknown constants $\nu$ and $\lambda$ is the same. Even so, for a given value of angular momentum $C$, we can obtain different values for $\lambda$ and $\nu$, which assures the stability of the circular orbit. The necessary and sufficient condition for the Lyapunov stability of the circular orbit in Manev potential is reduced to that from the two-body problem in Newtonian potential. Therefore we can conclude that the results from Manev two-body problem generalize those from the classical Newtonian case. The addition of the term inverse proportional with the third power of the distance between the bodies let us explain certain dynamical phenomena unexplained in the Newtonian gravitational field, but it does not destroy the Lyapunov stability of the circular orbit.

\end{document}